\documentclass[preprint,prab, nofootinbib]{revtex4-2}

\usepackage[colorlinks=true, allcolors=blue]{hyperref}
\usepackage{graphics}% Include figure files
\graphicspath{{images/}{}}
\usepackage{dcolumn}% Align table columns on decimal point
\usepackage{bm}% bold math
\usepackage{siunitx}
\usepackage{xcolor}
\usepackage{braket}
\usepackage{mathtools}
\usepackage{amsmath}
\usepackage{leftidx}
\usepackage{wasysym}
\usepackage{float}
\usepackage{cleveref}
\Crefname{equation}{Eq.}{Eqs.}

\usepackage{amsfonts} %% <- also included by amssymb
\DeclareMathSymbol{\shortminus}{\mathbin}{AMSa}{"39}

\begin{document}

%\pubblock

\title{Ion Coulomb Crystals in Storage Rings for
Quantum Information Science}

%\bigskip 
%\usepackage{authblk}
%\renewcommand\Authfont{\small}
%\renewcommand\Affilfont{\scriptsize}

%Author{K.~Brown}
%\Address{Brookhaven National Laboratory}

%\Author{S.~Nagaitsev\footnote{also at the University of Chicago.}}
%\Address{Fermilab, PO Box 500, Batavia, IL 60510-5011}
\author{S.~Brooks}
\author{K.~Brown\footnote{also at Stony Brook University}}
\author{F.~Méot} 
\author{A.~Nomerotski} 
\author{S.~Peggs} 
\author{M.~Palmer} 
\author{T.~Roser} 
\author{T.~Shaftan}
\affiliation{Brookhaven National Laboratory, Upton, NY 11973, USA}

%\author[2]{A.~Di~Meglio}
%\affil[2]{CERN openlab, 1211 Geneva 23, Switzerland}

%\author[3]{F.~Zimmermann}
%\affil[3]{CERN, 1211 Geneva 23, Switzerland}

\author{G.~H.~Hoffstaetter\footnote{also at Brookhaven National Laboratory}}
\affiliation{Cornell University, Ithaca, NY 14853, USA}

\author{S.~Nagaitsev\footnote{also at the University of Chicago}}
\author{J.~Lykken}
\author{J.~Jarvis} 
\author{V.~Lebedev} 
\author{G.~Stancari}
\author{A.~Valishev}
\affiliation{Fermilab, Batavia, IL 60510, USA}

%\author[6]{L.~Giannessi} 
%\affil[6]{INFN and Elettra-Sincrotrone, 34149 Basovizza, Trieste, Italy}

%\author[7]{L.~Civale}
\author{A.~Taylor} 
\author{A. Hurd}
\author{N.~Moody}
\affiliation{Los Alamos National Laboratory, Los Alamos, NM 87545, USA}

\author{P.~Muggli} 
\affiliation{Max-Planck-Institut für Physik, Munich 80805, Germany}

\author{A.~Aslam}
\author{S.~G.~Biedron\footnote{also with Element Aero} \footnote{also with the Center for Bright Beams, Cornell University}}
\author{T.~Bolin}
\author{S.~Sosa Guitron}
\affiliation{University of New Mexico, Albuquerque, NM 87131, USA}

\author{C.~Gonzalez-Zacarias}
\affiliation{University of Southern California, Los Angeles, CA 90007, USA}

\author{M.~Larsson}
\author{R.~Thomas}
\affiliation{Stockholm University, AlbaNova, Stockholm 10691, Sweden}

\author{B.~Huang}
\author{T.~Robertazzi}
\affiliation{Stony Brook University, Stony Brook, NY 11794, USA}

\author{J.~Cary} 
\affiliation{Tech-X Corporation, 5621 Arapahoe Avenue, Suite A, Boulder, CO 80303, USA}

\author{B.~M.~Hegelich}
\affiliation{University of Texas, Austin, TX 78712, USA}

\author{B.B.~Blinov}
\affiliation{University of Washington, Seattle, WA 98195, USA}

\author{S.~Milton}
\affiliation{Element Aero, Chicago, IL 60643, USA}
\date{\today}

\begin{abstract}
Quantum information science is a growing field that promises to take computing into a new age of higher
performance and larger scale computing as well as being capable of solving problems classical computers are incapable of solving. The outstanding issue in practical quantum computing today is scaling up the system while maintaining interconnectivity of the qubits and low error rates in qubit operations to be able to implement error correction and fault-tolerant operations. Trapped ion qubits offer long coherence times that allow error correction. However, error correction algorithms require large numbers of qubits to work properly. We can potentially create many thousands (or more) of qubits with long coherence states in a storage ring. For example, a circular radio-frequency quadrupole, which acts as a large circular ion trap and could enable larger scale quantum computing. Such a Storage Ring Quantum Computer (SRQC) would be a scalable and fault tolerant quantum information system, composed of qubits with very long coherence lifetimes. With computing demands potentially outpacing the supply of high-performance systems, quantum computing could bring innovation and scientific advances to particle physics and other DOE supported programs. Increased support of R$\&$D in large scale ion trap quantum computers would allow the timely exploration of this exciting new scalable quantum computer. The R$\&$D program could start immediately at existing facilities and would include the design and construction of a prototype SRQC. We invite feedback from and collaboration with the particle physics and quantum information science communities.
\end{abstract}

\maketitle

\def\thefootnote{\fnsymbol{footnote}}
\setcounter{footnote}{0}
\section*{Executive summary}
We call the attention of the particle physics and accelerator communities to the opportunity to advance and bring the scientific and technological expertise of accelerator science and technology to support advances in quantum information sciences (QIS).

The needs of particle physics for quantum computing systems are well documented and beyond the scope of this note. We will cite, however, the DOE HEP QuantiSED initiative to develop technology for quantum information science and quantum computing approaches for particle physics experiments~\cite{ref6}. A number of workshops were held to define the scientific needs and opportunities, including the HEP-ASCR Study Group Report, Grand Challenges at the Interface of Quantum Information Science, Particle Physics, and Computing from 2015~\cite{ref7}, the BES-HEP roundtable report, Common Problems in Condensed Matter and High Energy Physics from 2015~\cite{ref8}, the ASCR Report on Quantum Computing for Science from 2015~\cite{ref9}, and the HEP-ASCR QIS roundtable report, Quantum Sensors at the Intersections of Fundamental Science, QIS and Computing from 2016~\cite{ref10}. 

The particle physics quantum information science challenges include a need for greater computational power for simulations and reconstructing events. For example, researchers are working on a quadratic unconstrained binary optimization (QUBO)~\cite{ref11} for track reconstruction~\cite{ref12, ref12a, ref12b}. In a quantum implementation they treat a ground state of a Hamiltonian as a quantum machine instruction that is equivalent to a QUBO, easily transformed into an Ising model or Hopfield network. The same group is researching quantum associative memory, used with quantum algorithms for finding track candidates with a constant-time lookup. This is a commonly used real-time pattern recognition technique used in particle physics. We discuss these examples because they highlight the demanding computing needs for particle physics data analysis. The parallelism required and the ability to use quantum memory fit well with a storage ring quantum computer that potentially can contain 10’s of thousands of qubits that can act as computational elements or as memory elements.

Many accelerator laboratories around the world are forming groups to focus on exploring quantum computing as applied to particle physics, nuclear physics, applied sciences, and more. CERN Openlab has been partnering with corporations, other laboratories, and universities to advance in several computing topics, including quantum technologies~\cite{ref4}. In 2018 they held the Quantum Computing for High Energy Physics workshop, to focus on how to overcome significant challenges related to information and communications technologies in the next decade and beyond for the LHC and detector systems~\cite{ref5}. The challenges in bridging the computing needs with available computational resources were discussed in this workshop as well as ways that quantum computing can meet many of the challenges.

Trapped ions form isolated small quantum systems that have demonstrated low decoherence rates. Such systems can be controlled and measured using laser-induced manipulations of the ions. All the requirements needed to perform quantum simulation and quantum logic operations are met with such systems. The fundamental challenges, as is true for all quantum simulation systems, are in scaling and error correction. There are many other challenges, such as control-ability, cooling, constructing quantum circuits, and more. A storage ring quantum computing approach uses trapped ions but at a larger scale. Stationary ions may be in the same inertial reference frame as the lasers and other systems used to control and measure eigenstates, but a single set of lasers can only operate on a small number of ions in the trap. Very long ion chains would not be easily addressable~\cite{kranzl}. Moving ions, while in a different inertial reference frame, will pass by the system that control and measure the eigenstates. However, by overcoming the challenges, discussed later in this manuscript, large scale trapped moving-ion systems could be built and applied to the computing challenges in particle physics. 

\section{Technical overview}

 At BNL, efforts began in 2019 to study the use of crystalline beams in storage rings as platforms for quantum computing~\cite{ref1, ref1a,ref2, ref2a, ref2b}. At Fermilab, efforts have also begun to study the use of crystalline beams in storage rings~\cite{ref3}. At BNL and Fermilab there are programs for research in quantum information superconducting systems, materials for quantum information systems, and on quantum devices and sensors. At LANL, there is research on materials for quantum information, quantum devices and sensor and superconducting cavities.

A storage ring quantum computer will make use of a unique and proven particle accelerator device called a circular radiofrequency quadrupole (CRFQ). We can think of a CRFQ as an unbounded Paul trap, where ions circulate at some fixed velocity (and thus fixed frequency). Although the words ‘storage ring’ may evoke thoughts of a large device kilometers in circumference, a single CRFQ in our preliminary design is only one meter in circumference and can contain 10’s of thousands of ions that, when cooled sufficiently using laser cooling techniques, will form ion Coulomb crystals (ICC). Given that a single quantum computation involving a handful of logical error-corrected qubits makes use of thousands of physical qubits to implement an error-correcting code with fault tolerance, such a particle accelerator-based architecture enables high throughput, full quantum computations including the error correction overhead.

We can form two basic states of matter by cooling ion beams down to very low temperatures. The first is a classical crystalline beam, defined, to be “a cluster of circulating, charged particles in its classical lower-energy state subject to circumferentially varying guiding and focusing electro-magnetic forces and Coulomb interacting forces”. The second state of matter is an ultracold crystalline beam, or what we refer to as an ion Coulomb crystal, cooled well below the Doppler cooling limit, but not deeply into the Lamb-Dicke limit. In this regime the ion-laser interaction can be well approximated as the carrier transition that does not affect the motion (simple spin flip) and two sideband transitions, the red sideband and the blue sideband, that change the vibrational state by one quantum unit while flipping the spin. The higher-order interactions (i.e. 2nd, 3rd and higher-order sidebands) are strongly suppressed.
In this regime, thermal vibrations are small enough to distinguish the external quantum modes of the crystalline structure and to minimize any micromotion from the rf confining the ions in the center of the trap. [The laser cooling systems can reduce the motional degrees of freedom of the ions but cannot diminish kinetic energy from the induced micromotion.]

Ion trap systems exploit two quantum properties of the ions in the trap, external energy eigenstates, such as the axial center-of-mass motion of the string of ions in the trap, and the internal eigenstates of each ion in the string. When sufficiently cooled, the string of ions in the trap has properties that we can use to define a set of computational basis states operated on using laser excitations.

The primary method of establishing a qubit involves excitation and measurement of stable or metastable internal states of individual ions, such as the hyperfine states. The basic method, as described by Wineland et al., using $^9Be^+$ and then using the $^2S_{1/2}(F = 2, m_F = 2)$ and $^2S_{1/2}(F = 1, m_F = 1)$ hyperfine ground states (denoted $\ket{\downarrow}$ and $\ket{\uparrow}$, respectively), we can construct a practical qubit. Tuning a polarized laser beam to the $\ket{\downarrow} \rightarrow \; ^2P_{3/2}$ transition and by observing the scattered photons we can resolve two distinct spin states. With this technique, per Wineland et al., the quantum states can be determined with almost 100\% efficiency. Using Raman transitions and polarized light scattering we can ‘write’ and measure the hyperfine states.

\subsection{Laser-Ion Interactions}

In this section we descibe the fundamentals of how ions in ion traps are utilized as qubits, following the approach of C. Roos~\cite{roos}.
In ion traps, a qubit is coherently manipulated by exciting it with a monochromatic travelling-wave laser beam. This can be described with a Hamiltonian as,
\begin{eqnarray}
 H = H^e + H^m + H^i,
\end{eqnarray}
where $H^m$ describes the ion’s motion, $H^e$ the electronic structure of the ion, and $H^i$ is the interaction of the ion with the laser. For a laser-cooled ion in a harmonic potential, $H^m$ is the sum of three quantum harmonic oscillator Hamiltonians corresponding to the motional degrees of freedom.
\begin{eqnarray}
 H^m = \hbar \nu_{j} (a_j^\dagger a_j + 1/2),
\end{eqnarray}
where $\nu_{j}$, $j=x,y,z$, is the harmonic oscillator frequency, and $a_j^\dagger$ and $a_j$ are the corresponding creation and annihilation operators of the quantum vibrational eigenstates, $\ket{0}, \ket{1}, \ket{2}, ...$. The electronic structure of the ion can be approximated using a two-level model of the excited state $\ket{\uparrow}$ and ground state $\ket{\downarrow}$. Then:
\begin{eqnarray}
 H^e = \frac{\hbar \omega_0}{2} \sigma_z,
\end{eqnarray}
where $\omega_0$ is the atomic transition frequency, and $\sigma_z$ is the Pauli spin matrix. Finally, the laser-ion interaction Hamiltonian is,
\begin{eqnarray}
 H^i = \hbar \Omega (\sigma_+ + \sigma_-) \cos(k\hat{x} - \omega_L t - \phi_L).
\end{eqnarray}

\begin{figure}[htb]
\centering 
\includegraphics*[width=0.7\columnwidth]{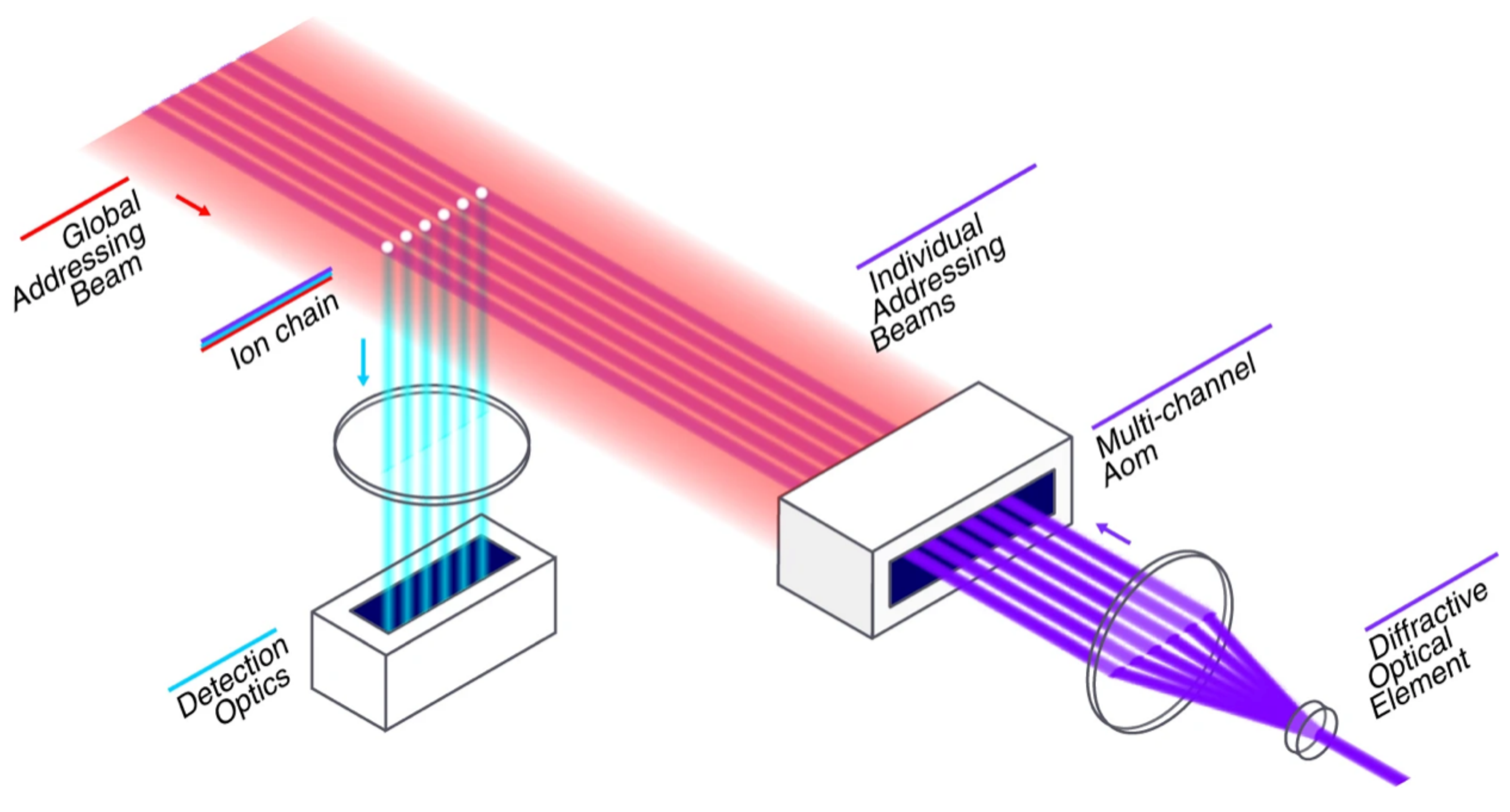}
\caption{From~\cite{wright}, a method to address individual atoms in an
ion trap could also be used to address multiple moving ions. This
is just to illustrate the concept, as more than the target ion(s) would
be illuminated in this scheme.}
\label{fig1}
\end{figure}

The frequency, $\omega_L$, and phase, $\phi_L$, of the laser field as seen by the ion, described by the position operator, $\hat{x}$, along with the laser wave vector, $k$, and Rabi frequency, $\Omega$ determine the strength of the interaction. The jump operators, $\sigma_+$ and $\sigma_-$ induce the transitions between qubit states $\ket{\downarrow}$ and $\ket{\uparrow}$. 

These interactions, whether with a stationary or moving ion, are unchanged. In the case of the moving ion, the time to perform a state transition is restricted to the laser width divided by the ions velocity, unless the laser field can be moved and follow in the same reference frame. One approach around this may be to spread the laser field out along the path of the ion, as shown in figure~\ref{fig1}.

\section{Technical advantages}

A significant challenge in quantum computing is controlling quantum decoherence. However, research in ion traps has shown that quantum states in trapped ions can persist for very long times, even on the scale of minutes. In a storage ring environment, we will have to be sure to eliminate any sources of noise and other forms of energy that may disrupt the trapped ion quantum states. There is a good discussion of this topic by Wineland, et al.~\cite{ref13}. The scaling of the number of qubits (N), while limited mostly to internal interactions, is related to the problem of decoherence. It has a unique temporal component we must consider, since we cannot operate on all ions in the crystalline beam simultaneously.

Inside the storage ring system, we can isolate groups of smaller numbers of ions from each other, using longitudinal rf potentials or by separating using velocity modulation in the cooling systems. We can then operate on these isolated sets of qubits independently. In the storage ring environment, there is potential for multiplexing as well as an ability to work on ions and groups of ions in parallel. A storage ring could contain thousands of these smaller individual crystals, producing 10’s of thousands of qubits. The many small chains of ions could serve different purposes, depending on the algorithm being employed. For example, we can use some ion chains as quantum memory and some for other purposes, such as for systematic or error analysis. Having many ions and ion chains available opens many possibilities; simultaneous computations, quantum memory, error correction, and more.

\section{Technical challenges}

While there are significant advantages to a storage ring approach compared to linear traps, the non-stationary ions present a number of technical challenges. Many details are outlined by Shaftan and Blinov~\cite{ref2, ref2a, ref2b}.

Cooling ions in a trap is well established~\cite{cool1, cool1a, cool1b}. Methods for 3D cooling, such as optical molasses or sympathetic cooling, are also well established~\cite{cool2,cool3}. However, cooling ions that are moving raises many challenges. Doppler laser cooling of moving ions was established in the PALLAS experiments~\cite{pallas, pallasa} and has been demonstrated at GSI using RF potentials as a counter force to a single laser system~\cite{gsi1}. The amount of detuning required for high Doppler shift raises a challenge for relativistic beams~\cite{gsi2, gsi3}. More research and new ideas are needed to find ways to perform 3D cooling of moving ions, to get to the temperatures needed to use these ions for QIS applications.

Addressing (setting and measuring eigenstates) moving ions poses new challenges to the systems used to irradiate the ions. Acoustic Optical Modulators (AOM) and Deflectors (AOD) are used to control the laser irradiation. Ions moving at sufficiently high velocities still require a fixed amount of time to irradiate. As seen in figure~\ref{fig1}, the laser pulses can be controlled with the AOM, but it may have to switch at very high frequencies. The amount of laser power and required switching times could exceed existing technology. Advances in AO devices would be an area of research that would benefit the storage ring approach.

Keeping track and managing large number of ions raises new technical challenges~\cite{huang}. Individual qubits need to be tracked, identifying and mitigating qubit loses needs to be done, and detecting potential ion reordering due to collisions with particles in the vacuum enclosure. This, along with scaling to large numbers of ions is an important area of research. The challenge is to collect the relevant information in as short a time as possible, to minimize latency in the QIS systems.

\section{Community building}

We believe a discussion and increased collaboration between the communities of particle physics, accelerator science, laser science, data science (including machine learning and high-performance computing)~\cite {AIforScience}, controls engineering, and quantum information science on methods to advance quantum computing to problems in particle physics and other disciplines will greatly advance the science and technology in these communities.

\section{Crossover Benefits}

The ability to create ultra-cold Coulomb crystals has interests beyond quantum computing applications. The idea of colliding crystalline beams has been discussed in the accelerator community for many years~\cite{wei}. This allows for the concept of high specific luminosity collisions to be explored~\cite{brooks}. Low temperature ion traps can also provide particles with a known wavefunction that could one day enable single-particle colliders and other exotic concepts.
The “specific luminosity”: luminosity per watt, or integrated
luminosity per joule of beam power, quantifies this energy efficiency.
Multiplying by kinetic energy gives the related measure “integrated
luminosity per particle”, which is $1.6\times 10^{-9}$ barn$^{-1}$ for a
single pass through an LHC interaction point (IP), and $2.8\times 10^{-5}$
barn$^{-1}$ for the ILC.  The ultimate limit is saturating the
cross-section, for example at 2 barn$^{-1}$ for a 500 mbarn ion
cross-section, which corresponds to every particle colliding.  This can
be done only if the de Broglie wavelength is short enough to localize
the particle transversely in a region with area smaller than the
cross-section.

Ultra-cold Coulomb crystals with high densities could also be of interest for researchers in astromaterials~\cite{caplan} and earth and planetary science.  The experimental study of such materials could benefit our understanding of, for example, neutron stars and other such exotic objects and associated materials. In earth and planetary science, simulating high compression effects using ICCs could be of interest. For example, there is a curious phenomenon of electrons (charge density) piling up between ion cores in certain crystals at high compression, and if this phenomenon is intrinsically quantum mechanical vs. being largely controlled by electrostatics albeit quantum modified ~\cite{Caldwell_1997} ~\cite{Ma_2009}.   Interestingly, ICCs and geomaterials would be subject to Earnshaw’s Theorem~\cite{Jones_1980} that no system of charged particles can be in static, stable equilibrium without external forces. ICC studies could also be compared to other studies carried out on classical colloidal crystals~\cite{Frenkel_2006, li2016} consisting of colloidal spheres in water and interacting by screened Coulomb repulsions.  Such classical, colloidal crystals exhibit overdamped dynamics owing to hydrodynamic interactions~\cite{hurd1985} yet may have thermodynamic kinship to ICCs.

\section{Experimental Test Beds}

Significant progress has been made in the ion trap community towards building practical QIS systems using stationary trapped ions~\cite{monroe}. Theoretical work on using a storage ring approach began only relatively recently, although studies of crystalline beams began with the seminal works by Schiffer and his collaborators~\cite{schiffer, schiffera, schifferb}. Important experimental studies of crystalline beams were performed at PALLAS and ASTRID rings~\cite{pallas,pallasa, astrid}. To further develop a storage ring QIS system both theoretical and experimental studies are necessary. The technological challenges need to be addressed with experimental tests and demonstrations.  Below we describe two facilities, where such experiments can begin immediately.

\subsection{IOTA}

\begin{figure}[htb]
\centering 
\includegraphics*[width=0.5\columnwidth]{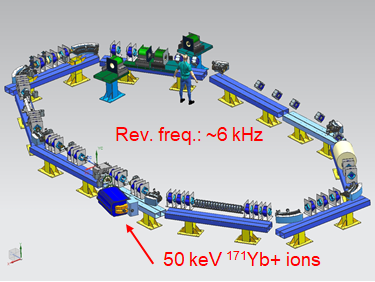}
\caption{Schematic of IOTA storage ring.}
\label{fig_IOTA}
\end{figure}

The Fermilab Integrable Optics Test Accelerator (IOTA) ring \cite{Antipov_2017} is a DOE-sponsored R\&D facility, focused on Accelerator and Beam Physics experiments. It is a 40-m circumference ring, capable of storing ion beams with momenta in the range of 70 - 150~MeV/c. It features a stretched-octagon layout with four 2-m long straight sections and two 6-m long sections allowing for the placement of essential beam manipulation and instrumentation devices as well as for the installation and easy replacement of experimental insertion devices.  Estimates show \cite{ref3} that a 50-keV (120-MeV/c momentum) $^{171}$Yb+ may be an ideal ion species for testing in the IOTA (see Fig.~\ref{fig_IOTA}). 

We propose to use this existing unique infrastructure and our expertise in beams and storage rings to create a 1-D crystal of ~one-million ions (~10 m with a 10-$\mu$m separation) as a pathfinder for scalability in Coulomb crystal-based SRQC techniques.  We aim at developing the full suite of capabilities and techniques to create these crystals, achieve ion temperatures of $<$50 $\mu$K, and prepare and readout quantum states using external lasers.  The proposed R\&D programn would include the following elements

\begin{itemize}
    \item \textbf{Beam cooling:} attain ion beams which are much colder than those available today and learn how to sustain their low temperatures in a steady state;
    \item \textbf{Advanced engineering:} develop techniques to reduce the amount of thermal and RF noise in storage rings;
    \item \textbf{Diagnostics:} develop compact and highly accurate diagnostics and control systems of the ion crystal's orbit around the macroscopic storage ring;
    \item \textbf{Quantum communications:} develop novel protocols for two-qubit and multi-qubit gates that involve ultra-fast pulsed lasers;
    \item \textbf{Optics:} design, develop and demonstrate a broadband optical modulator that is capable of handling high optical power of such lasers, while generating short pulses with high RF bandwidth; 
    \item \textbf{Computer modeling:} develop simulation codes that take into account radiation and absorption of photons, finite temperature of surrounding chambers, and vacuum conditions.
\end{itemize}

\subsection{DESIREE}
% notel this is just grabbed from the DESIREE website

\begin{figure}[htb]
\centering 
\includegraphics*[width=0.8\columnwidth]{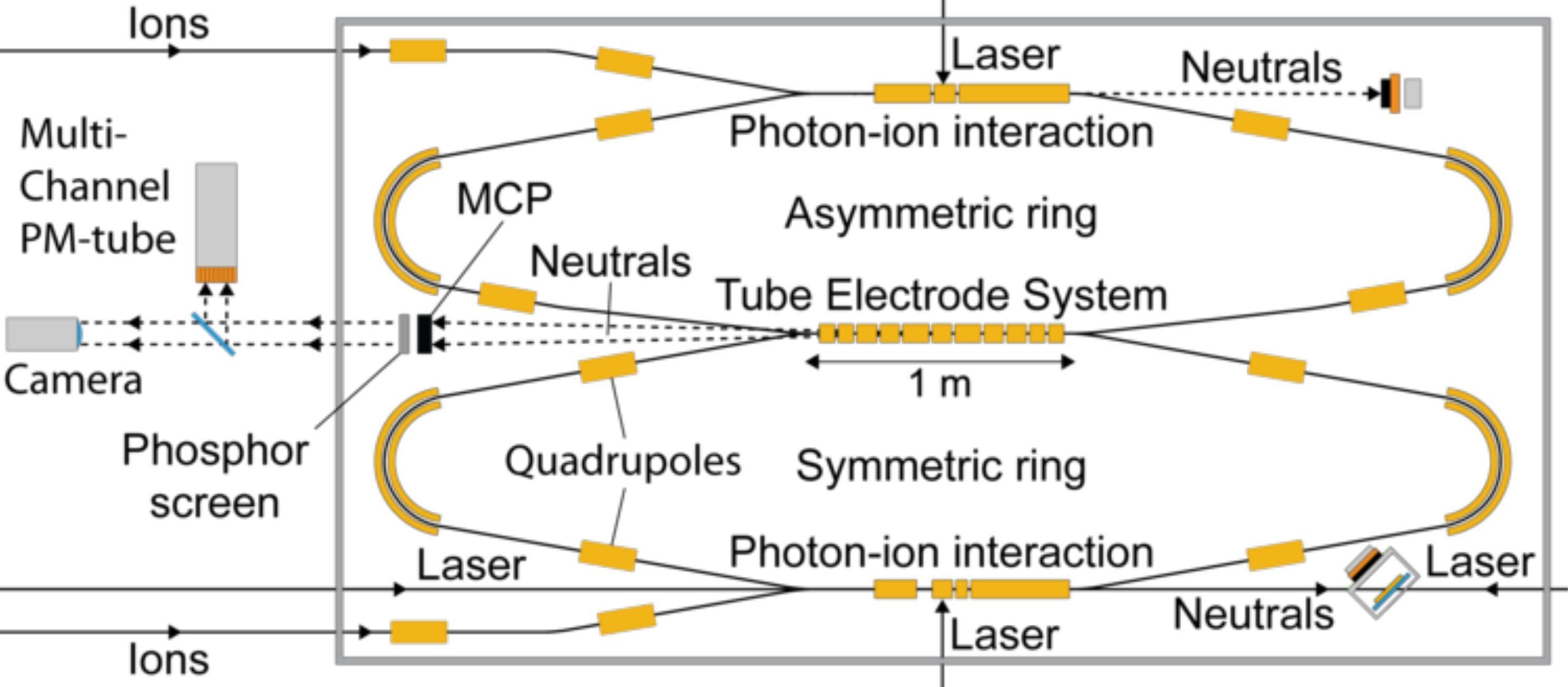}
\caption{Schematic of DESIREE electrostatic storage rings.}
\label{fig2}
\end{figure}

The DESIREE~\cite{desiree} storage rings, located at Stockholm University, have circumferences of 8.6 m and a common section for merging keV-beams of positive ions stored in one ring with keV-beams of negative ions stored in the other ring. In the merging section, both beams travel together and in the same direction through a common, biased, tube-electrode system (see figure~\ref{fig2}). The bias voltage can be set to control cation-anion collision energies down to the meV range. This allows studies of interactions between individual pairs of internally cooled ions (in their electronic, vibrational, and sometimes also rotational ground states) at collision energies (temperatures) similar to those in the interstellar medium.

The DESIREE storage rings offer a platform for studying moving ions and methods of cooling and addressing.

\section{Summary}

Applying accelerator expertise and technology to move quantum information systems to scales needed by particle physics applications means using accelerator science techniques and systems in new and creative ways to allow development of new kinds of quantum computing devices. We propose designing and building a storage ring quantum computer that follows a quantum gate-model architecture. This is a great opportunity to use the advanced methods developed in accelerator science to benefit quantum information science and eventually particle physics.

\section{Acknowledgements}

Biedron extends great thanks to Raymond Jeanloz of the University of California Berkeley for the useful discussions regarding the science of ICCs.

%%%%%%%%%%%%%%%%%%%%%%%%%%%%%%%%%%%%%%%%%%

%  If you would like to use BibTEX for the bibliography, please feel free to do so.  It is not required.

%  To use BibTeX,

%    1.  uncomment the following two lines, 
%    2.  comment out everything below from  \begin{thebibliography}{99}   to \end{thebibliography).
%    3.  create the file  myreferences.bib, and process this file in the usual way

\bibliographystyle{JHEP}
\bibliography{references}  % file references.bib

%%%%%%%%%%%%%%%%%%%%%%%%%%%%%%%%%%%%%%%%%

\end{document}